\documentclass[aps,showpacs,twocolumn,superscriptaddress]{revtex4}
\usepackage{amsmath}
\usepackage{amssymb}
\usepackage{amscd}
\usepackage{wasysym}
\usepackage[ansinew]{inputenc}
\usepackage[T1]{fontenc}
\usepackage{ae,aecompl}

    \usepackage[dvips]{graphicx}
    \DeclareGraphicsExtensions{.eps}

\newcommand{\be}{\begin{equation}}
\newcommand{\ee}{\end{equation}}
\newcommand{\bea}{\begin{eqnarray}}
\newcommand{\eea}{\end{eqnarray}}
\newcommand{\nn}{\nonumber}

\newcommand{\LL}{\mathcal{L}}

\newcommand{\tr}{{\rm tr}}

\begin{document}
\bibliographystyle{apsrev}

\title{Beyond mean-field dynamics in open Bose-Hubbard chains}
\author{D. Witthaut}
\email{witthaut@nld.ds.mpg.de}
\affiliation{Max-Planck-Institute for Dynamics and Self-Organization,
D--37073 G\"ottingen, Germany}
\author{F. Trimborn}
\affiliation{Institut f\"ur theoretische Physik, Leibniz Universit\"at Hannover,
D--30167 Hannover, Germany}
\author{H. Hennig}
\affiliation{Max-Planck-Institute for Dynamics and Self-Organization,
D--37073 G\"ottingen, Germany}
\author{G. Kordas}
\affiliation{Institut f\"ur theoretische Physik and Center for Quantum
Dynamics, Universit\"at Heidelberg, D--69120 Heidelberg, Germany}
\author{T. Geisel}
\affiliation{Max-Planck-Institute for Dynamics and Self-Organization,
D--37073 G\"ottingen, Germany}
\author{S. Wimberger}
\affiliation{Institut f\"ur theoretische Physik and Center for Quantum
Dynamics, Universit\"at Heidelberg, D--69120 Heidelberg, Germany}
\date{\today }

\begin{abstract}
We investigate the effects of phase noise and particle loss on the 
dynamics of a Bose-Einstein condensate in an optical lattice.
Starting from the many-body master equation, we discuss
the applicability of generalized mean-field approximations
in the presence of dissipation as well as methods to simulate quantum 
effects beyond mean-field by including higher-order correlation
functions.
It is shown that localized particle dissipation leads to surprising
dynamics, as it can {\em suppress} decay and {\em restore}
the coherence of a Bose-Einstein condensate. 
These effects can be applied to engineer coherent structures 
such as stable discrete breathers and dark solitons.
\end{abstract}

\maketitle

\section{Introduction}

Decoherence and dissipation, caused by the irreversible coupling of a 
quantum system to its environment, represent a major obstacle for the
long-time coherent control of quantum states. However, in the last years 
it has been realized that dissipation can be extremely useful if it can be 
controlled accurately. Recent experiments have shown that strong 
correlations can be induced by two-body losses in ultracold 
quantum gases \cite{Syas08,Garc09}. Three-body losses
can be tailored to generate effective three-body interactions \cite{Dale09} 
and to prepare strongly correlated states for quantum simulations of
color superfluidity \cite{Kant09}, quantum hall physics \cite{Ronc10}
or d-wave pairing \cite{Dieh10}.
Even more, dissipation can be used as a universal tool in quantum state 
preparation \cite{Diel08,Krau08}, entanglement generation \cite{Krau10}
and quantum information processing \cite{Vers08}.
These concepts of controlling quantum dynamics and transport are 
particularly important for experiments with ultracold atoms in optical 
lattices, where it is possible to address the quantum system with
single-site resolution \cite{Bakr09,Sher10}. An even higher spatial 
resolution has been realized with a focussed electron beam, removing
atoms one-by-one from the lattice \cite{Geri08,Wurt09}.
The effects of such a localized particle loss on the dynamics of 
a Bose-Einstein condensate (BEC) have been investigated from a nonlinear 
dynamics viewpoint in several papers in the last years, discussing 
the possibility to induce nonlinear structures such as bright breathers 
\cite{Livi06,Ng09}, dark solitons \cite{Braz09} or ratchets \cite{Carl06}.
These studies were based on a mean-field approximation, where 
the loss was introduced heuristically as an imaginary potential.

In this article we go beyond this approximation and investigate the 
quantum dynamics of ultracold atoms in a finite optical lattice with 
dissipation, which provides a distinguished model system for the 
study of open one-dimensional chains. 
Our analysis is based on a numerical integration of the full 
many-body master equation and generalized mean-field
methods. In section \ref{sec-method}, we present an explicit derivation 
of the mean-field equations of motion, which hold if the 
many-body state is close to a BEC, and generalize this  
approach to take into account higher order correlation 
functions \cite{Vard01,Tikh07}.
If particle loss is the only source of dissipation, the mean-field 
equations reduce to a non-hermitian Schr\"odinger equation
applied previously \cite{Ng09,Braz09}. While such a non-hermitian 
description has been thoroughly studied for single particle quantum 
mechanics \cite{Fyod05}, the applicability to open many-body 
systems is an open issue.

Two important cases are 
studied in detail: In section \ref{sec-boundary}, we analyze how 
boundary dissipation induces localization and purifies a  BEC. 
In section \ref{sec-local}, we consider localized 
loss from a single lattice site, which creates a vacancy and leads to 
a fragmentation of the condensate. Remarkably, strong dissipation 
can suppress the decay of the condensate and a coherent dark 
soliton can be generated by properly engineering the dynamics.
The techniques presented here can be directly applied in ongoing 
experiments \cite{Geri08,Wurt09}.

\section{The mean-field limit and beyond}
\label{sec-method}

The coherent dynamics of ultracold atoms in optical lattices
is described by the celebrated Bose-Hubbard Hamiltonian \cite{Jaks98}
\bea
  \hat H = - J  \sum \nolimits_{j} \left( \hat a_{j+1}^{\dagger} \hat a_j +
                  \hat a_{j}^{\dagger} \hat a_{j+1} \right)
         + \frac{U}{2} \sum \nolimits_j 
           \hat a_{j}^{\dagger}  \hat a_{j}^{\dagger} 
               \hat a_{j}  \hat a_{j},
    \label{eqn-hami-bh}
\eea
where $\hat a_j$ and $\hat a_j^\dagger$ are the bosonic annihilation and 
creation operators in mode $j$, $J$ denotes the tunneling
matrix element between the wells and $U$ the interaction strength. 
We set $\hbar = 1$, thus measuring energy in frequency units.
This model assumes that the lattice is sufficiently deep, such that
the dynamics takes place in the lowest Bloch band only.

In the presence of dissipation, the dynamics is given by a master
equation in Lindblad form \cite{Breu02},
\be 
  \dot{\hat \rho} =  -i [\hat H,\hat \rho]  + \LL \hat \rho.
  \label{eqn-master}
\ee
Here, we are especially interested in the effects of
localized particle loss, which can be implemented by an 
electron beam \cite{Geri08,Wurt09} or by a strongly focussed resonant 
blast laser. Furthermore, phase noise is always present in
experiments, which degrades the phase coherence between 
adjacent wells and heats the sample 
\cite{Gati06,08phase2}.
These two processes are described by the Liouvillians
\cite{Breu02,Angl97,Ruos98,Pich10} 
\be
  \LL_{\rm loss} \hat \rho = -\frac{1}{2} \sum \nolimits_j \gamma_j \left(
     \hat a_{j}^{\dagger} \hat a_{j}  \hat \rho
     + \hat \rho \hat a_{j}^{\dagger} \hat a_{j} 
     - 2 \hat a_j \hat \rho \hat a_{j}^{\dagger}  \right),
\ee
\be     
    \LL_{\rm phase} \hat \rho = 
    -\frac{\kappa}{2} \sum \nolimits_j \hat n_j^2 \hat \rho + \hat \rho \hat n_j^2
       - 2 \hat n_j \hat \rho \hat n_j,
\ee
where $\gamma_j$ denotes the loss rate at site $j$ and
$\kappa$ is the strength of the phase noise.

To derive the mean-field approximation, we start from 
the  single particle  reduced density matrix (SPDM)
$\sigma_{jk} = \langle \hat a_j^\dagger \hat a_k \rangle
=\tr(\hat a_j^\dagger \hat a_k \hat \rho)$ 
\cite{Vard01,Tikh07,08mfdecay}.
The equations of motion for $\sigma_{jk}$ are  obtained from the 
master equation (\ref{eqn-master}),
\bea  
  i\frac{d}{dt} \sigma_{j,k} 
  &=&  \tr \left(\hat a_j^\dagger \hat a_k [\hat H, \hat \rho]
                  + i \hat a_j^\dagger \hat a_k   \LL \hat \rho \right) \nn \\
  &=&   -J \left( \sigma_{j,k+1} + \sigma_{j,k-1}
        - \sigma_{j+1,k} - \sigma_{j-1,k} \right) \nn \\
  && + U  \left( \sigma_{kk} \sigma_{jk} + \Delta_{jkkk} 
               - \sigma_{jj} \sigma_{jk} - \Delta_{jjjk} \right),  \nn \\ 
   &&   - i \frac{\gamma_j + \gamma_k}{2} \sigma_{j,k} 
     - i \kappa (1-\delta_{j,k}) \sigma_{j,k} ,
               \label{eqn-mf-final}
\eea
where we have defined the covariances
\be
   \Delta_{jk \ell m} = 
      \langle \hat a_{j}^\dagger \hat a_{k}  \hat a_{\ell}^\dagger \hat a_{m} \rangle
       - \langle \hat a_{j}^\dagger \hat a_{k} \rangle 
        \langle \hat a_{\ell}^\dagger \hat a_{m} \rangle .
      \label{eqn-cov}  
\ee
In the mean-field limit $N \rightarrow \infty$ with $g = UN$ fixed,
one can neglect the variances $\Delta_{jk \ell m}$ in 
Eq.~(\ref{eqn-mf-final}) in order to obtain a closed set of 
evolution equations. This is appropriate for a pure BEC, because 
the variances scale only linearly with the particle number $N$, 
while the products $\sigma_{jk} \sigma_{\ell m}$ scale as $N^2$.
If phase noise can be neglected, i.e. $\kappa = 0$, the equations
of motion (\ref{eqn-mf-final}) are equivalent to the non-hermitian 
discrete nonlinear Schr\"odinger equation 
\be
  i \frac{d}{dt} \psi_k = -J (\psi_{k+1} + \psi_{k-1}) 
 + U |\psi_k|^2 \psi_k - i \frac{\gamma_k}{2} \psi_k
    \label{eqn-dnlse}
\ee
by the identification $\sigma_{j,k} = \psi_j^* \psi_k$. 
This provides a proper derivation of the non-hermitian 
Schr\"odinger equation, which has previously
been applied heuristically \cite{Livi06,Ng09,Braz09}.

The mean-field approximation assumes a pure
BEC and is strictly valid only in the
limit $N \rightarrow \infty$. To describe many-body 
effects such as quantum correlations and the depletion of
the condensate for large, but finite particle numbers, we 
generalize the Bogoliubov backreaction  (BBR) method
\cite{Vard01} to the dissipative case, taking into account 
the covariances (\ref{eqn-cov}) explicitly.
We start with the coherent part of the master equation, which 
yields the following evolution equations for the four-point 
functions:
\bea
  &&  \!\!\!\!   i\frac{d}{dt}  \langle \hat a_{j}^\dagger \hat a_{m}  
         \hat a_{k}^\dagger \hat a_{n} \rangle
    =      \tr \left(\hat a_{j}^\dagger \hat a_{m}  
         \hat a_{k}^\dagger \hat a_{n} [\hat H, \hat \rho] \right) \nn \\   
   &&  = (\epsilon_m + \epsilon_n - \epsilon_j - \epsilon_k) 
       \langle \hat a_{j}^\dagger \hat a_{m}  
         \hat a_{k}^\dagger \hat a_{n} \rangle \nn \\
  && \; - J \, \langle \hat a_{j}^\dagger \hat a_{m}  
         \hat a_{k}^\dagger \hat a_{n+1}  + 
          \hat a_{j}^\dagger \hat a_{m}  
         \hat a_{k}^\dagger \hat a_{n-1} 
         + \hat a_{j}^\dagger \hat a_{m+1}  
         \hat a_{k}^\dagger \hat a_{n}    \nn \\
   && \qquad +  \hat a_{j}^\dagger \hat a_{m-1}  
         \hat a_{k}^\dagger \hat a_{n}  - 
          \hat a_{j+1}^\dagger \hat a_{m}  
         \hat a_{k}^\dagger \hat a_{n} 
         - \hat a_{j-1}^\dagger \hat a_{m+1}  
         \hat a_{k}^\dagger \hat a_{n}     \nn \\
     &&  \qquad -  \hat a_{j}^\dagger \hat a_{m}  
         \hat a_{k+1}^\dagger \hat a_{n} 
       -  \hat a_{j}^\dagger \hat a_{m}  
         \hat a_{k-1}^\dagger \hat a_{n} \rangle  \nn \\
   && \; + U \, \langle  \hat a_j^\dagger \hat a_m \hat n_m  
           \hat a_k^\dagger \hat a_n    +   \hat a_j^\dagger \hat a_m
           \hat a_k^\dagger \hat a_n \hat n_n  \nn \\
   && \qquad \quad -  \hat n_j \hat a_j^\dagger \hat a_m  
           \hat a_k^\dagger \hat a_n    -   \hat a_j^\dagger \hat a_m
           \hat n_k \hat a_k^\dagger \hat a_n  \rangle.
   \label{eqn-eom4-raw}  
\eea
Again, the interaction hamiltonian leads to higher-order
correlation functions. To obtain a closed set of
evolution equations, these function are truncated according to 
\cite{Tikh07}
\bea
  &&   \langle  \hat a_j^\dagger \hat a_m  \hat a_k^\dagger \hat a_n 
       \hat a_r^\dagger \hat a_s \rangle \approx
    \langle  \hat a_j^\dagger \hat a_m  \hat a_k^\dagger \hat a_n \rangle
    \langle \hat a_r^\dagger \hat a_s \rangle \nn \\
  && \qquad \qquad  +  
     \langle  \hat a_j^\dagger \hat a_m  \hat a_r^\dagger \hat a_s \rangle
    \langle \hat a_k^\dagger \hat a_n \rangle
   +  \langle \hat a_k^\dagger \hat a_n \hat a_r^\dagger \hat a_s \rangle
       \langle \hat a_j^\dagger \hat a_m \rangle \nn \\
 && \qquad \qquad    - 2     \langle  \hat a_j^\dagger  \hat a_m  \rangle
        \langle  \hat a_k^\dagger \hat a_n \rangle
       \langle \hat a_r^\dagger \hat a_s \rangle.
\eea
For a BEC, the six-point function scale as $N^3$, while the
error introduced by this approximation increases only linearly 
with $N$.
The relative error induced by the truncation thus vanishes 
as $1/N^2$ with increasing particle number.
Close to a pure condensate, the BBR method thus provides
a better description of the many-body dynamics than the 
simple mean-field approximation, since it includes the dynamics
of higher order methods at least approximately.
Using this truncation, the coherent part of the dynamics 
is given by
\bea
  &&  \!\!\!\!   i\frac{d}{dt} \Delta_{jmkn} 
    =  \nn \\
  && \!\!\! - J \big[ \Delta_{j,m,k,n+1} +    \Delta_{j,m,k,n-1}
                +  \Delta_{j,m+1,k,n} +    \Delta_{j,m-1,k,n} \nn \\
   && \;   -  \Delta_{j,m,k+1,n} -  \Delta_{j,m,k-1,n}
                -  \Delta_{j+1,m,k,n} -  \Delta_{j-1,m,k,n} \big] \nn \\
   && \!\!\! + U  \big[ \Delta_{mmkn}  \sigma_{jm} - \Delta_{jjkn}  \sigma_{jm}
           + \Delta_{jmnn}   \sigma_{kn}  -  \Delta_{jmkk} \sigma_{kn} \nn \\
   && \; +  \Delta_{jmkn} \left( \sigma_{mm} + \sigma_{nn}
                               - \sigma_{kk}  - \sigma_{jj} \right) \big].  
   \label{eqn-eom4-final}  
\eea

Particle loss and dissipation affect the dynamics of the 
four-point functions as follows
\bea
     && \frac{d}{dt} \langle \hat a_{j}^\dagger \hat a_{m}  
         \hat a_{k}^\dagger \hat a_{n} \rangle
       = \tr \left[ \hat a_{j}^\dagger \hat a_{m}  
         \hat a_{k}^\dagger \hat a_{n}   \LL \hat \rho    \right]   \nn \\
      && \quad = - \frac{\gamma_j + \gamma_m + \gamma_k + \gamma_n}{2} 
                           \langle \hat a_{j}^\dagger \hat a_{m}  
         \hat a_{k}^\dagger \hat a_{n} \rangle
          - \delta_{mk} \gamma_m \langle \hat a_j^\dagger \hat a_n \rangle \nn \\
        && \qquad   - \kappa  \, (2 + \delta_{mn} + \delta_{jk} - \delta_{jm} - \delta_{jn} 
        - \delta_{km} - \delta_{kn})  \nn \\
        && \qquad \quad \times   \langle \hat a_{j}^\dagger \hat a_{m}  
         \hat a_{k}^\dagger \hat a_{n} \rangle   . \nn
\eea
In terms of the variances this yields
\bea
    && \frac{d}{dt} \Delta_{jmkn}
       =  - \frac{\gamma_j + \gamma_m + \gamma_k + \gamma_n}{2} 
       \Delta_{jmkn} - \delta_{mk} \gamma_m \sigma_{jn} \nn \\
      && \quad  - \kappa ( \delta_{mn} + \delta_{jk} - \delta_{jn}  - 2 \delta_{km})
       (\Delta_{jmkn} + \sigma_{jm} \sigma_{kn}) \nn \\
       && \quad - \kappa ( 2 - \delta_{jm} - \delta_{kn} ) \Delta_{jmkn}.
\eea

The BBR method is especially useful if the many-body state is
close to, but not exactly equal to a pure BEC. In particular, it 
accurately predicts the onset of the depletion of the condensate
mode. The number of atoms in this mode is given by the 
leading eigenvalue $\lambda_0$ of the SPDM $\sigma_{j,k}$,
where the trace of $\sigma_{j,k}$ gives the total number of 
atoms $n_{\rm tot}$. The ratio $\lambda_0/n_{\rm tot}$
is referred to as the condensate fraction \cite{Vard01,Legg01}.

\begin{figure}[bt]
\centering
\includegraphics[width=8.2cm,  angle=0]{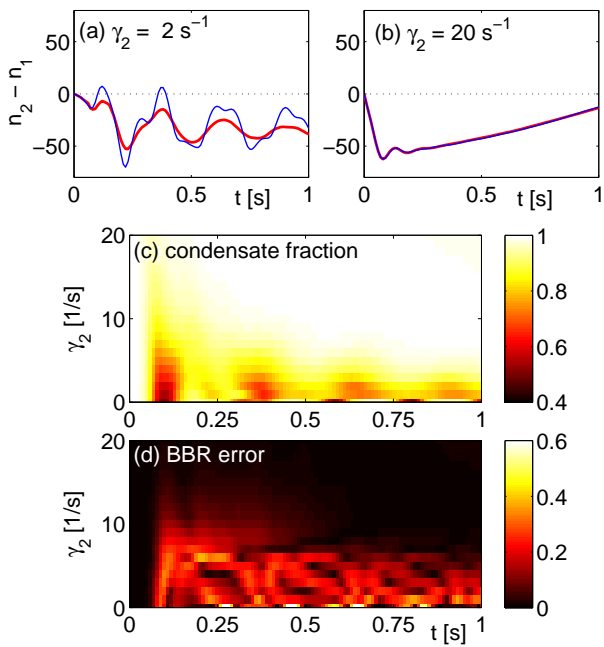}
\caption{\label{fig-bbrtest}  
(Color online)
Numerical test of the BBR methods for a leaky double-well trap
with loss in the second well. (a,b) Dynamics of the population
imbalance $\langle \hat n_2 - \hat n_1 \rangle$ for two different
values of the loss rate, comparing the BBR approximation 
(solid blue line) to numerically exact results (thick red line).
(c) Condensate fraction $\lambda_0/n_{\rm tot}$ as a function of time
and the loss rate $\gamma_2$.
(d) Trace distance (\ref{eqn-tdist}) between the exact rescaled 
SPDM $\sigma(t)/n(t)$ and the respective BBR approximation.
In all cases the initial state is assumed to be a pure BEC
with with equal population and a phase difference of $\pi$ 
between the two modes. The remaining parameters are
$J = 10 \, {\rm s}^{-1}$, $\kappa=0$, $U = 0.5 \, {\rm s}^{-1}$ 
and $n(0) = 200$ atoms.
}
\end{figure}

The BBR approach has been extensively tested for closed systems
in \cite{Tikh07}. Therefore, we only briefly comment on the performance
of this method in the presence of dissipation.
Figure \ref{fig-bbrtest} shows two examples of the dynamics of a BEC in a
leaky double-well trap, comparing the BBR approximation 
(solid blue line) and numerically exact results (thick red line).
The initial state is assumed to be a pure BEC with equal
population and a phase difference of $\pi$ between the two 
modes. In the case of strong dissipation, the BBR approximation
predicts the correct evolution of the population imbalance 
$\langle \hat n_2 - \hat n_1 \rangle$ with an astonishing
precision. In contrast, significant differences are observed 
for weak losses. This means that the presence of particle loss
actually improves the performance of the BBR method, as
the dissipation drives the many-body quantum state towards
a pure BEC \cite{08stores}. This is confirmed by the numerical
results for the condensate fraction $\lambda_0/n_{\rm tot}$ plotted in
Fig.~\ref{fig-bbrtest} (c). A significant depletion of the condensate 
is only observed for small values of the loss rate $\gamma_2$.
For a further quantitative analysis of the accuracy, we compare 
exact and BBR results for the rescaled SPDM $\sigma(t)/n_{\rm tot}(t)$. 
Figure \ref{fig-bbrtest} (d) shows the trace distance of the exact 
matrix and the matrix obtained by the BBR method,
\be
  d := \frac{1}{2} \tr \big( | \sigma_{\rm BBR}/n_{\rm BBR} - 
              \sigma_{\rm ex}/n_{\rm ex} | \big),
              \label{eqn-tdist}
\ee
as a function of time for different values of $\gamma_2$.
For sufficiently large dissipation, one observes that the
distance approximately vanishes for all times. In this
regime the quantum dynamics is faithfully reproduced
by the BBR approximation.

\section{Boundary dissipation}
\label{sec-boundary}

\begin{figure}[bt]
\centering
\includegraphics[width=8.2cm,  angle=0]{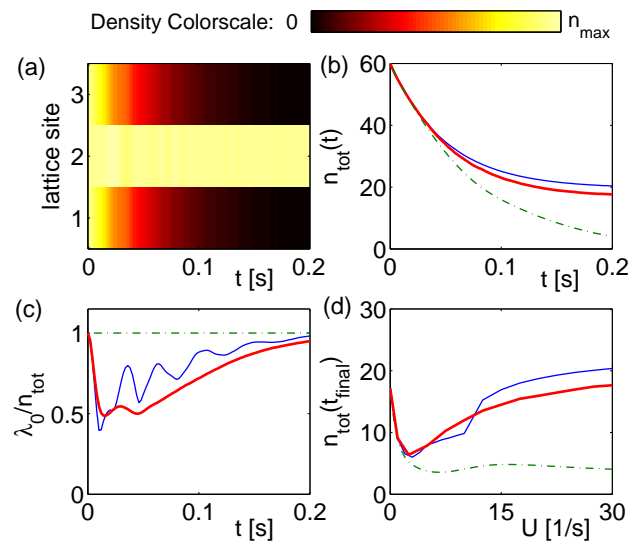}
\caption{\label{fig-breather}  
(Color online)
Dynamics of a BEC in a triple-well trap with boundary dissipation:
(a) atomic density $\langle \hat n_j(t) \rangle$,
(b) total particle number and (c) the condensate fraction
$\lambda_0/n_{\rm tot}$ for $J = 5 \, {\rm s}^{-1}$, 
$\gamma = 20 \, {\rm s}^{-1}$, $\kappa=0$, 
$U = 30 \, {\rm s}^{-1}$ and $n(0) = 60$ atoms.
(d) Total particle number after a fixed propagation time 
$t_{\rm final} = 0.2 \, {\rm s}$ as  a function of the interaction
strength $U$. 
Mean-field ($- \cdot -$) and BBR (---) results are compared to 
numerically exact simulations with a quantum jump method 
averaging over 200 trajectories (thick red line). 
}
\end{figure}

We first analyze the effects of boundary dissipation with a focus on
small systems for which numerically exact solutions of the 
many-particle dynamics are still possible, for instance
by the quantum jump method \cite{Dali92,Breu02}. 
A comparison to numerically exact results for these
examples provides another test of performance of
the BBR approach.

We consider the decay of an initially pure, homogeneous BEC in a 
triple-well trap with boundary dissipation. 
Figure~\ref{fig-breather} (a) and (b) show the evolution of the atomic 
density and the total particle number for strong inter-atomic
interactions $U = 30 \, {\rm s}^{-1}$. One observes a fast decay of
the atoms at the outer sites  while the population at the 
central site is remarkably stable. This is confirmed by the evolution of 
the total particle number, which rapidly drops to about one third of its 
initial value, where it saturates for a long time. This is a consequence 
of the dynamical formation of a discrete breather at the central site, 
which is an important generic feature of nonlinear lattices. Generally, 
discrete breathers, also called discrete solitons, are spatially localized, 
time-periodic, stable excitations in perfectly periodic discrete systems \cite{Trom00,Camp04,Flac08,Henn10}. They  arise intrinsically from 
the combination of nonlinearity and the discreteness of the system.
In the presence of boundary dissipation, these excitations become 
attractively stable such that the quantum state of the atoms will 
converge to a pure BEC with a breather-like density for a wide 
class of initial states. Once a discrete breather is formed, it remains 
stable also if the dissipation is switched off. The crucial role of strong 
interactions is illustrated in Fig.~\ref{fig-breather} (d), where the residual 
atom number after $t_{\rm final} = 0.2 \, {\rm s}$ of propagation is plotted 
as a function of the interaction strength.  The particle number increases for 
large values of $U$ to $n_{\rm tot}(t_{\rm final}) \approx 20$
due to the breather formation.

For strong interactions a simple mean-field 
approximation fails. It strongly underestimates the 
residual particle number as it predicts that discrete breather are formed
only for stronger losses. In contrast, the BBR 
results agree well with the many-particle simulation even for 
large values of $U$. We thus conclude that quantum
fluctuations facilitate the formation of repulsively bound structures.
Furthermore, a mean-field approach cannot account for genuine 
many-body features of the dynamics. Figure \ref{fig-breather} (c) 
shows the evolution of the condensate fraction $\lambda_0/n_{\rm tot}$, 
where $\lambda_0$ is the leading eigenvalue of the SPDM \cite{Legg01}. 
In the beginning, interactions lead to a rapid depletion of the
condensate. On a longer time scale, however, 
dissipation restores the coherence and drives the atoms to a pure BEC 
localized at the central lattice site \cite{08stores}.
The BBR approach faithfully reproduces the depletion and 
re-purification  but additionally predicts 
unphysical temporal revivals. This example thus demonstrates
the strength but also the limitations of this method. 

\begin{figure}[bt]
\centering
\includegraphics[width=7cm,  angle=0]{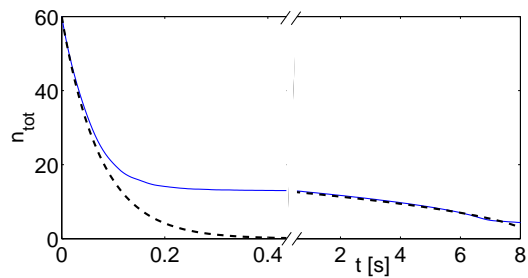}
\caption{\label{fig-bdecay}  
(Color online)
Decay of a discrete breather state for $J = 5 \, {\rm s}^{-1}$, 
$\gamma = 20 \, {\rm s}^{-1}$, $\kappa=0$ and $U=10 \, {\rm s}^{-1}$.
Numerical results calculated with the BBR method (---) are
compared to the analytic estimates (\ref{eqn-bdappr1}) and 
(\ref{eqn-bdappr2}), respectively ($- \, - \, -$).
}
\end{figure}

The decay dynamics of the discrete breather state is further
analyzed in Fig.~\ref{fig-bdecay}. The total atom number 
$n_{\rm tot}(t)$ decreases rapidly until the discrete breather 
is formed at $t \approx 0.2 {\rm s}$. Afterwards the decay is
much slower and clearly non-exponential. 
In both regimes, one can calculate the evolution of 
$n_{\rm tot}(t)$ approximately, starting from the relation
$\dot n_{\rm tot} = - \gamma (n_1 + n_3)$.
Initially, all sites are filled homogeneously, 
$n_1 = n_3 = n_{\rm tot}/3$,
such that the total particle number decays as
\be
  n_{\rm tot}(t) \approx n_{\rm tot}(0) e^{-2\gamma/3 \, t }.
  \label{eqn-bdappr1}
\ee
When the discrete breather is formed, the population of the outer 
wells is given by $n_1=n_3 = J^2/(U^2 n_{\rm tot})$ in first
order perturbation theory. The atom number then decays as
\be
  n_{\rm tot}(t) \approx \sqrt{n_{\rm db}^2 - 4 \gamma J^2 \, t /U^2 },
  \label{eqn-bdappr2}
\ee
where $n_{\rm db}$ is the number of atoms bound in the discrete
breather state. 
Both approximations are compared to the BBR simulation 
results in Fig.~\ref{fig-bdecay}, assuming a breather with
$n_{\rm db} = 13$ atoms. One observes an excellent agreement 
in the both regimes, i.e. an exponential decay for very short times 
($t \apprle 0.1 \, {\rm s}$) and an algebraic decay when the discrete 
breather is formed. The transition between the linear and 
nonlinear decay takes place at $t \approx 0.2 \, {\rm s}$.
A deviation from the algebraic decay (\ref{eqn-bdappr2}) for the
discrete breather is observed only for very long times when the
atom number is very small such that the simple perturbative 
estimate for $n_{1,3}$ is no longer valid.

\section{Localized loss}
\label{sec-local}

Recent experiments with ultracold atoms have demonstrated an
enormous progress in spatial addressability 
using specialized optical imaging systems \cite{Bakr09,Sher10}
or a focussed electron beam \cite{Geri08,Wurt09}.
Especially the latter experiment allows to manipulate a 
Bose-Einstein condensate in an optical lattice dissipatively  
with single-site resolution. In the following, we
study the quantum dynamics in a finite lattice of 11 sites with 
closed boundary conditions and loss occurring from the central
site only, which leads to remarkably different decay as in the 
case of boundary dissipation studied above.

\begin{figure}[tb]
\centering
\includegraphics[width=8.4cm,  angle=0]{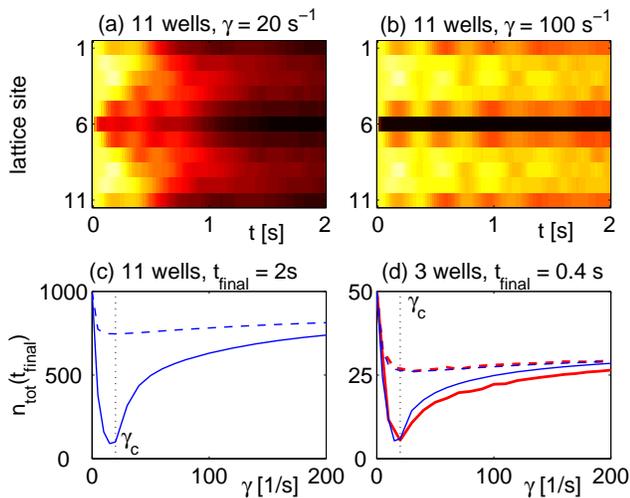}
\caption{\label{fig-dsoliton1}
(Color online)
Formation of vacancies by localized loss at the central
lattice site.
(a,b) Evolution of the atomic density $\langle \hat n_j(t) \rangle$ 
(colorscale as in Fig.~\ref{fig-breather}).
(c,d) Final value of the total particle number after a fixed 
propagation time $t_{\rm final}$ as a function of
the loss rate $\gamma$, without (solid lines) and
with strong phase noise (dashed lines, $\kappa = 50 \, {\rm s}^{-1}$).
The remaining parameters are $J = 5 \, {\rm s}^{-1}$, 
$U = 0.2 \, {\rm s}^{-1}$ and $n(0) = 1000$ particles
(a-c) and $U  = 2 \, {\rm s}^{-1}$ and $n(0) = 50$ particles (d).
The dynamics has been simulated with the BBR (thin blue lines) 
and the quantum jump method (thick red lines). 
}
\end{figure}

A remarkable feature of the quantum dynamics is illustrated in 
Figure~\ref{fig-dsoliton1}, showing the results of a BBR 
simulation for an initially pure homogeneous BEC.
For a modest loss rate $\gamma = 20 \, {\rm s}^{-1}$, atoms tunnel 
to the central site where they are dissipated with a rate $\gamma$, 
such that  the BEC decays almost homogeneously.
On the contrary, stronger losses ($\gamma = 100 \, {\rm s}^{-1}$) lead 
to a formation of a stable vacancy. The central site is rapidly depleted, 
but the atoms in the remaining wells are mostly unaffected.
Thus one faces the paradoxical situation that an increase of the loss
rate can suppress the decay of the BEC. Two effects contribute
to this counterintuitive behavior: (i) The absorbing potential suppresses
tunneling to the leaky lattice site. This effect is present
also in the linear case and can be explained by an analogy to
wave optics \cite{Syas08}: A large mismatch of the index of refraction
leads to an almost complete reflection of a wave from a surface.
This is true for an imaginary index describing an absorption 
as well as for a real index. 
(ii) A dark breather stabilizes the vacancy and prevents the 
flow of atoms to the central site. This nonlinear structure
 remains stable also if the dissipation 
is reduced or switched off afterwards 
(cf. \cite{Camp04,Flac08,Henn10}
for a discussion of the stability of breathers).

The suppressed decay of the BEC is further illustrated in 
Fig.~\ref{fig-dsoliton1} (c,d), where the residual atom number 
after a fixed propagation time is plotted as a function of the 
loss rate $\gamma$. The coherent output of the system, 
i.e.~the number of lost atoms, assumes a maximum for a 
finite loss rate $\gamma_{c}$. This maximum is reminiscent of the
quantum stochastic resonance discussed in \cite{08stores}.
In the following we will estimate the value of $\gamma_{c}$ by 
determining a lower bound for $\gamma$ for the dynamical 
breather formation. 
As a single (both bright and dark) breather exhibits a 
pronounced population imbalance between the central site 
and the neighboring sites, we estimate $\gamma_{c}$ by 
matching the timescales of dissipation $\tau_{D}=2/\gamma$
and tunneling $\tau_{J}$, i.e., $\tau_{D}=\tau_{J}$. For smaller
values of $\gamma$, atoms can tunnel away from the leaky lattice
site again before they are lost, while for larger values of $\gamma$ 
a population imbalance can form.
From Eq.~(\ref{eqn-dnlse}) we read $\tau_{J}=1/(2J)$ where 
the factor $1/2$ accounts for atoms tunneling from two sites to 
the leaky site. Hence, the critical loss rate is estimated as 
$\gamma_{c}=4J$. We find good agreement of our qualitative 
estimate for $\gamma_{c}$ (dotted vertical lines in 
Fig.~\ref{fig-dsoliton1} (c,d)) with the dip in the total particle 
number. 
An important quantity for the breather formation and stability 
is the effective nonlinearity of the system $\lambda=U n_{tot}(t)/2J$, 
which, due to particle loss, is time-dependent. Strikingly, 
though $\lambda$ depends on the interaction strength $U$ 
(which is different in Fig.~\ref{fig-dsoliton1} (c) and (d)), the 
fairly good estimate $\gamma_{c}$  is independent of $U$.

Figure \ref{fig-dsoliton1} (d) shows the respective results for a 
triple-well trap with loss from the central site. A comparison of the 
BBR approximation to a numerically exact many-particle simulation
shows a good agreement for all values of $\gamma$.
Phase noise suppresses decay as it  effectively decouples 
the lattice sites. Thus, only the atoms initially loaded
at the leaky lattice site decay as $e^{-\gamma t}$,
while the other atoms remain at their initial positions.
With increasing loss rate $\gamma$, the number of atoms 
lost from the trap approaches $\approx n(0)/M$ as shown
in Fig.~\ref{fig-dsoliton1} (c,d). 

\begin{figure}[tb]
\centering
\includegraphics[width=8.2cm,  angle=0]{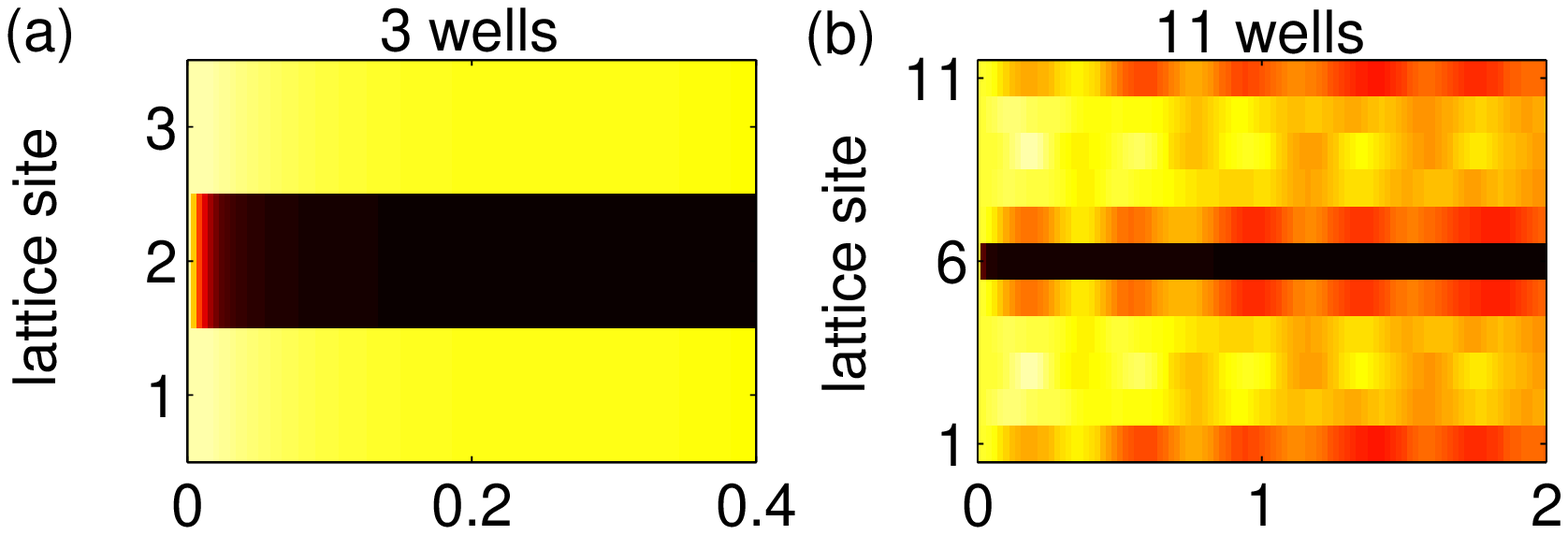}
\includegraphics[width=8.2cm,  angle=0]{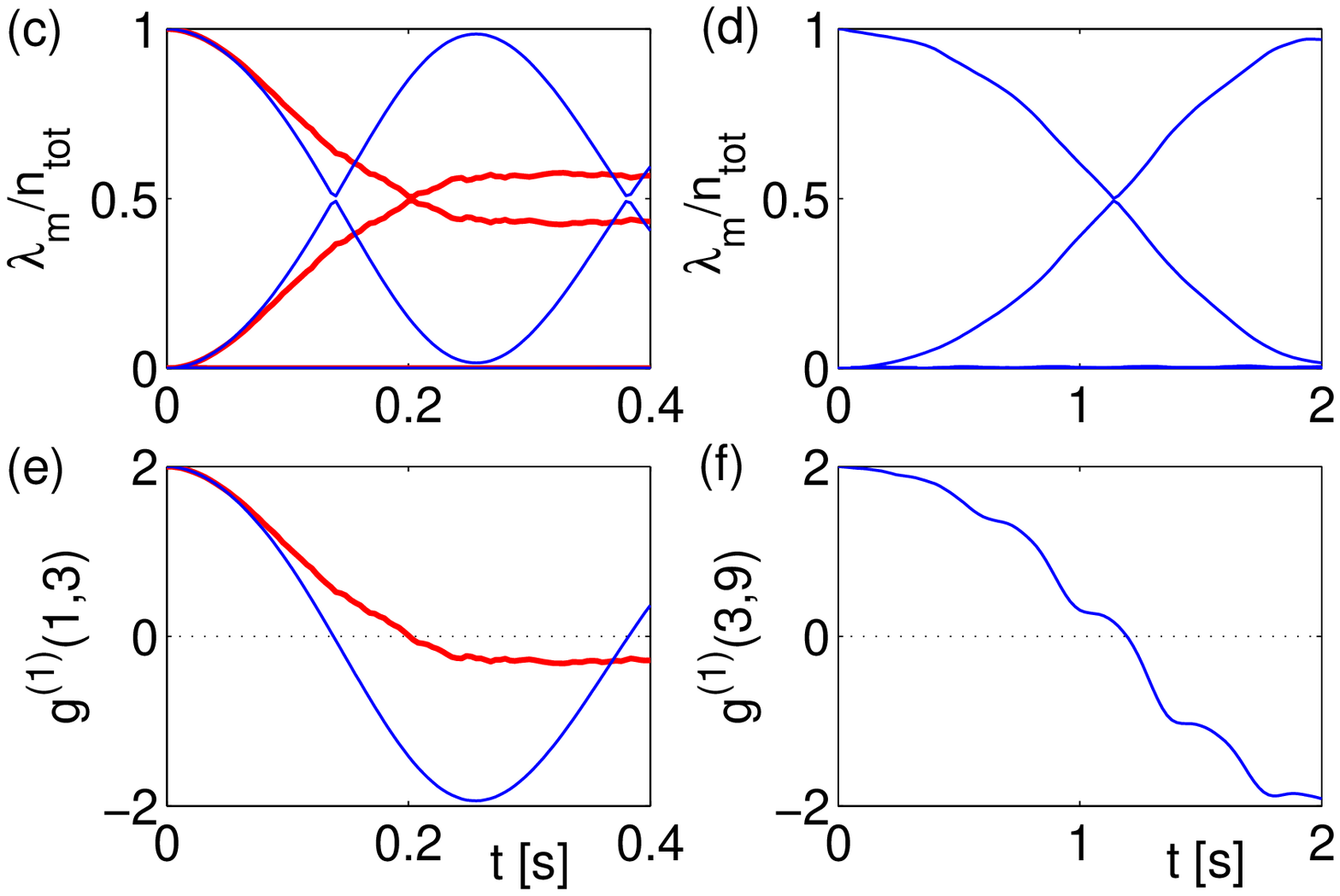}
\caption{\label{fig-dsoliton-coh}
(Color online)
Coherence of a vacancy generated by loss from the central site:
(a,b) atomic density, (c,d)
scaled eigenvalues $\lambda_m/n_{\rm tot}$ of the SPDM 
and (e,f) phase coherence $g^{(1)}$ between the two 
BEC fragments. Parameters are the same as in 
Fig.~\ref{fig-dsoliton1}  with $\gamma = 100 \, {\rm s}^{-1}$
and $\kappa = 0$.
Results of a quantum jump simulation are plotted as 
thick red lines, BBR results as thin blue lines.
}
\end{figure}

The previous reasoning suggests to use dissipation as a tool
to coherently engineer the quantum state of a BEC in an optical 
lattice. Mean-field theory predicts that dissipation can be used to 
efficiently create a coherent dark soliton \cite{Braz09}, but cannot 
assert the coherence of the final state as discussed above.
The results of a BBR and a quantum jump simulation of the 
many-body dynamics shown in  Fig.~\ref{fig-dsoliton-coh}
reveal the limitations of the phase coherence of a soliton
generated by local dissipation. 
The upper panels (a,b) show the rescaled eigenvalues 
$\lambda_m/n_{\rm tot}$ of the SPDM \cite{Legg01}.
One observes that there are two macroscopic eigenvalues
approaching 1/2, while all remaining eigenvalues vanish 
approximately. This proves that the dissipation generates 
a fragmented BEC consisting of two incoherent parts rather 
than a single BEC with a solitonic wavefunction. 
The BBR simulations correctly describe the fragmentation
of the condensate, but predict temporal revivals of the 
coherence which must be considered as artifacts of the 
approximation.
Experimentally, one can test the coherence by the interference
of the two fragments in a time-of-flight measurement. 
Figure~\ref{fig-dsoliton-coh} (e,f) shows the coherence
\be
   g^{(1)}(\ell,m) = \frac{\langle \hat a_\ell^\dagger \hat a_m
          +  a_m^\dagger \hat a_\ell \rangle}{
      \sqrt{\langle \hat n_\ell \rangle \langle \hat n_m \rangle}}
\ee
between the wells $\ell$ and $m$. 
One clearly observes the breakdown of phase coherence
between the two condensate fragments.

\begin{figure}[tb]
\centering
\includegraphics[width=8.2cm,  angle=0]{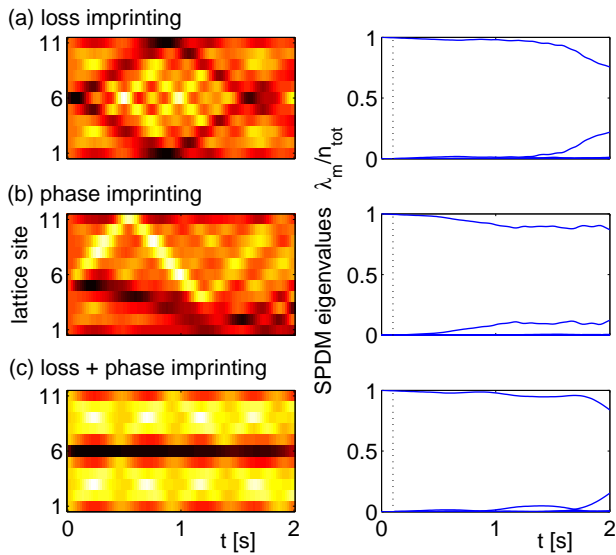}
\caption{\label{fig-dsol-switch}
(Color online)
Generation of dark solitons using loss imprinting at a rate
$\gamma = 100 \, {\rm s}^{-1}$ at the central
site (a,c) and phase imprinting in the lower half of the lattice (b,c),
both for times $t < 0.1 \, {\rm s}$  only. Shown are the atomic  density
(left, colorscale as in Fig.~\ref{fig-breather}) and the
scaled eigenvalues of the SPDM (right)
calculated with the BBR method.
Parameters are $J = 5 \, {\rm s}^{-1}$, 
$U = 0.1 \, {\rm s}^{-1}$, $\kappa=0$ and 
$n(0) = 1000$ particles.
}
\end{figure}

In order to overcome the loss of coherence, one can, however, engineer
the many-body dynamics. Figure \ref{fig-dsol-switch} illustrates
the generation of dark solitons comparing three different strategies. 
If the dissipation is switched off after the generation of a vacancy
at $t = 0.1 \, {\rm s}$, the condensate remains pure for long times.
However, the vacancy is not stable but decays into two dark
solitons traveling outwards \cite{Braz09}, where they are reflected 
at the boundaries.
The effects of a phase imprinting, which is an established
experimental method \cite{Dens00}, are shown in 
Fig.~\ref{fig-dsol-switch} (b).
A local potential is applied to the lower half of the lattice 
for $t < 0.1 \, {\rm s}$ imprinting a phase difference 
of $\pi$. Again coherence is preserved but the generated solitons
travel outwards. 
A coherent and stable dark soliton can be engineered by
combining both methods, as shown in Fig.~\ref{fig-dsol-switch} (c).
The generated dark soliton stays at its initial position
and remains coherent over a long time.

\section{Conclusion}

We have discussed the influence of localized particle dissipation 
on the dynamics of a finite one-dimensional Bose-Hubbard chain,
which describes a Bose-Einstein condensate in a deep optical 
lattice \cite{Geri08,Wurt09}. 
Starting from the many-body master equation, we have derived
the mean-field approximation and the dissipative Bogoliubov 
backreaction method, which allows a consistent calculation of 
the depletion of the condensate.

Two important special cases have been studied in detail.
Particle loss at the \emph{boundary} leads to localization and the 
formation of coherent discrete breathers. Surprisingly, 
dissipation together with interactions can re-purify a 
BEC.
A striking effect of \emph{localized loss} is that strong dissipation
can effectively {\it suppress} decay and induce 
stable vacancies. The decay shows a pronounced maximum for 
intermediate values of the loss rate, when the timescales 
of the dissipation and the tunneling are matched. 
Combined with an external potential, these effects can 
be used to generate stable coherent dark solitons.
These examples show that  engineering the dissipation is a promising
approach for controlling the dynamics in complex quantum 
many-body systems.

Ultracold atoms provide a distinguished model system for the
dynamics of interacting quantum systems, such that the effects 
discussed in the present paper may be observed in different 
systems, too. In particular, quantum transport of single excitations 
driven by \emph{local} dissipation has recently been studied in a 
variety of physical systems ranging from spin chains \cite{Clar10}
to light-harvesting biomolecules  \cite{Saro10}.
On the other hand, it has also been shown on the mean-field level
that nonlinear excitations such as discrete breathers play an 
important role for quantum transport in these systems 
(cf.~\cite{Camp04,Flac08,Zolo01} and references therein). 
Thus it is of general interest to further explore the regime which
interpolates between the nonlinear mean-field dynamics and the
many-body quantum dynamics in the spirit of the work presented
here.

\begin{acknowledgements}

We thank T.~Pohl for helpful comments and M.~K. Oberthaler for 
stimulating discussions and ideas on experimental possibilities.
We acknowledge financial support by the Deutsche Forschungsgemeinschaft 
(DFG) via the Forschergruppe 760 (grant number WI 3426/3-1), the Heidelberg 
Graduate School of Fundamental Physics (grant number GSC 129/1)
and the research fellowship programme (grant number WI 3415/1-1), 
as well as the Studienstiftung des deutschen Volkes.

\end{acknowledgements}

% --- Literatur -------------------------------------------------------------------

\end{document}